\documentstyle[11pt]{article}  
 
\begin{document}

\begin{center}
 
{\bf On the Properties of Konishi-Kaneko Map}

\vspace{0.2in}
 
V.G.~Gurzadyan, A.~Melkonyan, and K.Oganessyan
 
\vspace{0.1in} 

Astronomy Centre, University of Sussex, UK and 
Department of Theoretical Physics, Yerevan Physics Institute, Yerevan
375036, Armenia.
\end{center}

\vspace{0.1in}


\vspace{0.1in} 

An interesting possibility of study of the dynamics of $N$-body 
gravitating systems 
is provided by the Konishi-Kaneko iterated map [\ref{KonK}-\ref{KIn}]. 
Particular interest has the demonstrated
by Inagaki [\ref{Inag}] agreement of Konishi-Kaneko map with
the thermodynamic considerations and the results of numerical simulations.

At least two points outline the importance of iterated maps:

1. Iterated maps can enable one to avoid in certain way
the principal difficulties associated with $N$-body systems -- the 
non-compactness of the phase space and singularity of Newtonian 
interaction.

2. Iterated maps can be rather informative in revealing the
mechanisms of developing of chaos.

The Konishi-Kaneko map is defined as follows 
[\ref{KonK},\ref{KonKErg}]:
\begin{equation}
\label{R1}
p^{n+1}_i=p^n_i + k\sum_{j=1}^{N}\sin 2\pi(x^n_j-x^n_i),
\end{equation}
\begin{equation}
\label{R2} 
x^{n+1}_i=x^n_i+p^{n+1}_i;  (mod1) .
\end{equation}
This system describes 1-dimensional $N$-body system with a potential of 
interaction which is free of singularity and is attractive if $k > 0$. 
It represents a symplectic 2D-map of interval $(0,1)$ and
should have some common properties with maps known few decades
ago, such as the 
Ulam map [\ref{Ulam}]
proposed to describe the Fermi mechanism of acceleration of
cosmic rays, and the map of Zaslavsky and Sinakh 
[\ref{Zasl}]. 
Among the recent interesting studies in this area we mention the
papers by Kim (see e.g.[\ref{Kim}]), where the symmetrically coupled two 1D 
systems are studied and the existence of Feigenbaum bifurcations is shown,
with the value of Feigenbaum 
constant $\delta=8.72..$.  
Chaotic properties of the present non-symmentrical, i.e. Konishi-Kaneko 
system were observed in
[\ref{KonKErg}], where the Lyapunov numbers of
the system were calculated for clustered and non-clustered states.
Note, that at $N=2$ we have an integrable system.

The Jacobian of Konishi-Kaneko system is:
$$
\frac{\partial(y^{n+1}, x^{n+1})}{\partial(y^n, x^n)}=1.
$$
The corresponding Hamiltonian system was studied in 
[\ref{Inag}] with respect the thermodynamic instability. 

Obviously not every map defined on $(0,1)$ interval can possess Feigenbaum
bifurcations. Therefore 
first we have to check the necessary condition of existing of period-doubling
bifurcations, i.e. the negativity of the Schwartzian derivative:
\begin{equation}
Sf\equiv f'''/f''-3/2(f''/f')^2 < 0.
\end{equation}
This condition is fulfilled for Eqs.(\ref{R1}),(\ref{R2}) since $f'''/f' < 0$
for any value of $k$.

To obtain the bifurcation scale ${\delta}$ we have to find out the values of
period-doubling bifurcation points which must satisfy the conditions 
$$ 
\sum_{j}^{N} |x_{j}^{n+1} - x_{j}^n| < \epsilon, (2^1=2);\\ 
\sum_{j}^{N} |x_{j}^{n+2} - x_j^n| < \epsilon, \sum_{j}^{N} |x_j^{n+3} - 
x_j^{n+1}| < \epsilon, (2^2=4);\\
$$
for each $k_n, n=1,2,...$, respectively.
The ${\epsilon}$ is the accuracy of the obtained values of $k_n$. 
The accuracy of calculation of $k_1$ e.g. for $N=10$ was $\epsilon\approx 
10^{-5}$, and  $10^{-4}$ for $k_2$ and $k_3$. These calculations were
enough to find out the Feigenbaum universal number $\delta=8.72...$.
The results of calculations for $N=10$ are given in Table 1.
\begin{center}
\newpage

Table 1

\begin{tabular}{|l|l|l|l|l|l|}
\hline
\   $k_2-k_1 $ & $k_3-k_2$ & $\delta $\\
\hline
 $0.000196\dots $ & $0.00002222\dots $  & $8.82\dots$\\
\hline
 $0.000194\dots$ & $0.00002222\dots$ & $8.78\dots$ \\
\hline
 $0.00019368\dots$ & $0.00002222\dots$ & $8.71584\dots$\\
\hline
 $0.00019368\dots$ & $0.00002210\dots$  & $8.76359\dots$\\
\hline
 $0.00019368\dots$ & $0.00002225\dots$ & $8.70490\dots$\\
\hline
 $0.00019368\dots$ & $0.00002223\dots$ & $8.71219\dots$\\
\hline
 $0.00019368\dots$ & $0.00002221\dots$ & $8.71950\dots$\\
\hline
 $0.00019368\dots$ & $0.00002219\dots$ & $8.72682\dots$\\
\hline
 $0.00019368\dots$ & $0.000022205\dots$ & $8.72315\dots$\\
\hline
 $0.00019368\dots$ & $0.000022209\dots$ & $8.71950\dots$\\
\hline
 $0.00019368\dots$ & $0.000022207\dots$ & $8.72315\dots$\\
\hline
\end{tabular}
\end{center}

The results e.g. for $N = 3,5,7$ were absolutely identical 
with those for $N=10$, though with different accuracy $\epsilon$. We have
noticed a clear decrease in the accuracy with the increase of the
number of particles. 

The estimation of the values of $k_n$ requires careful 
procedure of calculations because of the complicated character of the 
system  and of the sensitivity on the iterations of $k_n$ and the
accuracy $\epsilon$.

Using the obtained values of $k_n$ and the formula
\begin{equation}
k_{\infty} = {(\delta k_{n+1} - k_n)} / {(\delta-1)},
\end{equation}
we also estimate the $k_{\infty}$, from which the chaotic behavior 
of the system is established and the map never repeats itself:
$$
k_{\infty} = 0.1307\dots .
$$
At $k>k_{\infty}$ the system should have positive Lyapunov numbers
as shown  in [\ref{KonKErg}].

The period-doubling points correspond to the phase transitions of second
order [\ref{AO}],
and can enable  the study of such systems via the methods of
thermodynamic formalism [\ref{BT}].

\end{document}